\documentstyle[prd,aps]{revtex}
\begin{document}
\textheight=640pt
\textwidth=440pt
\topmargin=-25pt
\leftmargin=-25pt
\hoffset -0.4cm

\title{A new hierarchy of avalanches observed in Bak-Sneppen evolution model}
\author{W. Li\thanks{E-mail: liw@iopp.ccnu.edu.cn} 
    and X. Cai\thanks{E-mail: xcai@wuhan.cngb.com} \\ 
   \footnotesize \sl Institute of Particle Physics, Hua-zhong Normal
                     University, Wuhan, 430079, P. R. China}

\maketitle
\vskip 0.5cm
\begin{abstract}
A new quantity, $\bar f$, denoting the average fitness of the ecosystem, 
is introduced in Bak-Sneppen model. Through this new quantity, a new hierarchy 
of avalanches, $\bar f_0$-avalanche, is observed in the evolution of 
Bak-Sneppen model. An exact gap equation governing the self-organization 
of the model in terms of $\bar f$ is presented. It is found that 
self-organized threshold $\bar f_c$ of $\bar f$ can be exactly obtained. 
Two basic exponents of the new avalanche, $\tau$, avalanche distribution, 
and D, avalanche dimension are given through simulations of one- and two
dimensional Bak-Sneppen models. It is suggested that $\bar f$ may be a good
quantity in determining the emergence of criticality.    
\end{abstract} 
\vskip 0.2cm

\mbox{} \hspace{1.225cm}PACS number(s): 87.10.+e, 05.40.+j, 64.60.Lx

\twocolumn
\hoffset 0.5cm
\rightskip -0.2cm
The term of avalanche may originate from the phenomena occurred in nature.
It is referred to as a sequential events which may cause devastating
catastrophe. The phenomena of avalanches are ubiquitous in nature. The
canonical example of avalanche in nature is the mountain slide, during which
great mass of snow and ice at a high altitude slide down a mountain side,
often carrying with it thousands of tons of rock, and sometimes destroy
forests, houses, etc in its path [1]. Since avalanches occur everywhere,
from the sandpile or ricepile, to the Himalayan sandpiles; from the
river network, to the earthquake, starquakes and even solar flares; from the
biology, to the economy, [2], etc, it is hence proposed [2] that avalanches 
may be the underlying mechanism of the formation of various geographical
structures and complex organisms, e.g., brains, etc. Furthermore, avalanches
may be the origin of fractals in the world. From this point of view,
avalanches can be viewed as the immediate results of complex systems, and 
hence can be used as the theoretical justification for catastrophism. This
is because if the real world is complex then the catastrophes are inevitable
and unavoidable in biology, history and economics. It is now even
proposed by Meng et al [3] that the formation of colorless gluon clusters
may be attributed to avalanches intrigued by the emission or absorption of
gluons.

Plenty of patterns provided by nature exhibit coherent macroscopic
structures developed at various scales and do not exhibit elementary
interconnections. They immediately suggest seeking a compact description of
the spatio-temporal dynamics based on the relationship among macroscopic
elements rather than lingering on their inner structure [4]. That is, one
needs to condense information when dealing with complex systems. Maybe only
this way is efficient and turns out successful.

As known, avalanche is one kind of macroscopic phenomenon driven by local
interactions. The size of avalanche, spatial and temporal as well, may be
sensitive to the initial configuration, or more generally, the detailed
dynamics of the system. However, the distribution of avalanches,
Gutenberg-Ritcher law [5], or equivalently, power-law, does not depend on such
kind of details due to the universality of complexity. Hence, in this sense
avalanche study may be an appropriate tools in studying various complex
phenomena. On the other hand, observation of a great variety of patterns,
such as self-similar, fractal behavior in nature [6,7,8,9], $\frac {1} {f}$
noise in quasar [10], river flow [11] and brain activity [12], and many
natural and social phenomena, including earthquakes, economic activity and
biological evolution suggests that these phenomena are signatures of
spatiotemporal complexity and can be related via scaling relations to the
fractal properties of the avalanches [13]. This suggests the occurrence of 
these general, empirical phenomena may be attributed to the same underlying
avalanche dynamics. Thus, one can see that study of avalanche is crucial in
investigating the critical features of complex systems. It can be even
inferred that avalanche dynamics does provide much useful information for us
to understand the general features of the ubiquitous complexity around us.
That is probably also why this paper focuses on such kind of topic.

Despite the fact that avalanche may provide insight into complexity, the
definition of which can be vastly different for various systems, and the
same kinds of systems, even the same system. Let us recall some definitions
of avalanches given before. In sandpile model [2], an avalanche is intrigued
by adding a grain or several grains of sand into the system at some time 
and causing the topple of some sites, which may later on cause some other
sites to topple. The avalanche is considered over when the height of all
sites are less than the critical value, say, 4. In Bak-Sneppen model [14],
several kinds of avalanche [13] are presented. For instance, $f_0$-avalanche, 
G(s)-avalanche, forward avalanche, backward avalanche, etc. Despite the fact
that these kinds of definitions of avalanches may show the various
hierarchical structures they manifest the same underlying fractal feature of
the ecosystem, i.e., self-organized criticality. Relating all these kinds of
avalanches one can provide a general definition of the avalanche for Bak-Sneppen
model: An avalanche corresponds to sequential mutations below certain
threshold. One can see that this kind of definition can ensure the mutation
events within a single avalanche are casually and spatially connected. In
addition, with this definition there exists a hierarchy of avalanches, each
defined by their respective threshold. It is the hierarchical structure of
the avalanche that exhibits the fractal geometry of the system and that
implies complexity.                  

It can be inferred from the definition of avalanche that there always exists a
triggering event which initiates the avalanche and whose effect, that is,
causing an avalanche to spread within the system later and later on, will
disappear at the end of the avalanche. And, the observation of avalanche
through the triggering event, up to now, is based on the individual level,
despite that the avalanche is a macroscopic and global phenomenon of the
system studied, in the laboratory, and in nature as well. Take sandpile
model, the triggering event is adding a grain or several grains of sand to
some sites and causing them to topple, thus initiating an avalanche.
Consider another model, Bak-sneppen model, in which the corresponding
triggering event of an avalanche is mutation of the extremal species causing
the fitness [14] of the extremal site at the next time step less than a
certain threshold. One can see that in the above two models triggering
events are directly concerned with the feature of individuals, e.g., the
height of the site in the former model, or, the fitness of the extremal site
in the latter one. It can be readily learned that the triggering events,
whether those in the laboratory or those in nature, are not directly related
to the global feature of the systems although avalanche can span across the
whole systems. Generally speaking, the observation of avalanches is done
through some feature of individuals, instead of that of the system as a
whole. However, general feature of the complex system may provide insight
into knowing the tendency of the evolution of the system. Specifically,
global feature of a complex system may help one to understand the critical
behavior of the system. That is, it is feasible that some characteristic
quantities representing the corresponding global features can be employed in
describing the critical behavior of the system. Furthermore, these
quantities ought to be related to avalanche dynamics, and hence can be used
to describe complexity emerged in a variety of complex systems. Apparently,
our aim is to search for or define such kind of quantities and then to
expect to observe new kind of avalanche based on these quantities. Indeed,
we obtain a new quantity which can be used to define a new hierarchy of
avalanches in Bak-Sneppen model. We suggest that this quantity may be used
as a criterion in determining the emergence of criticality. It will be shown
later that this new kind of avalanche still exhibits spatio-temporal
complexity in a different context.

Consider Bak-Sneppen model [14], which is a very simple evolution model of
biological ecosystem. Despite the simplicity of the model itself, it can
exhibit the skeleton of species evolution, punctuated equilibrium behavior.
Detailed information about this original model of evolution can be
available in Ref. [14]. In Bak-Sneppen model, each species is represented by a
single fitness. The fitness may represent population of a whole species or
living capability of the species [15]. Hence, one can see that fitness is a
vital quantity and is the only one describing the model. No other additional 
quantities are considered in this oversimplified model. Thus, the fitness is
the most important feature of species and that of the model. So, when
considering the global feature of the species ecosystem, one has to relate
this general feature to the feature of individuals. That is, the general
feature of the ecosystem should be associated with the fitness of the
species. As previously mentioned, a corresponding quantity should be found
to describe this general feature. Before presenting such kind of quantity
let us briefly review Bak-Sneppen model so that the readers who are not
so familiar with this model can have a rough idea of what it is about. 

Bak-Sneppen model is perhaps the simplest model of self-organized
criticality. In this "toy" model, random numbers, $f_i$, chosen from a flat
distribution, $p(f)$, are assigned independently to each species located on
a d-dimensional lattice of linear size $L$. At each time step, the extremal
site, i.e., the species with the smallest random number, together with its
$2d$ nearest neighboring sites, is chosen for updating by assigning $2d+1$
new random numbers also chosen from the same uniform distribution $p(f)$ to them.
This updating process continues indefinitely. After a long transient process
the system reaches a statistically stationary state where the density of
random numbers in the system vanishes for $f < f_c$ and is uniform above
$f_c$, the self-organized threshold. 

Having briefly introduced the model, next, we will introduce a new quantity.
Please note that the model we used is still Bak-Sneppen model. We observe
the evolution of the model without adding anything to the model. We simply
introduce the new quantity based on the fitness of the species. Define the
average fitness, denoted by $\bar f$, as below,
\begin{equation}
\bar f=\frac {1} {L^d} \sum \limits_{i=1}^{L^d} f_i
\end{equation}
\noindent
, where $f_i$ is the fitness of the $ith$ species. Here, we refer to $\bar
f$ as the average fitness of the whole system and as a global quantity.
$\bar f$ may represent the average population or average living capability
of the whole ecosystem. Large $\bar f$, i.e., high average fitness, may imply the
total population of the system is immense or its average living capability
is great, and vice versa. Initial value of $\bar f$, denoted by $\bar f (0)$
, can be easily calculated. As known, at the beginning of the evolution $f_i$'s are   
uniformly distributed between (0,1). So, for an infinite system, $\bar f (0)$
equals to 0.5. However, for a finite-size system $\bar f (0)$ will fluctuate 
slightly due to the finite
size of the system, which is not so important in the latter evolution. We
will simply consider the average value, 0.5. It should be pointed out that
$\bar f (0)$ does not reflect the correlation among species. As the
evolution goes on the correlation among the species within a system will
become more and more distinctive. Denote $\bar f (s)$ the average fitness of
the system at time step $s$ in the evolution. Hence, in the $s$ limit, i.e.,
$s \gg L^d$, $\bar f (s)$ may partly reflect information about correlation.
As a global quantity, $\bar f (s)$ should include information concerning the
interaction between species. Hence, it is natural to expect that $\bar f$
may be a good quantity in describing the feature of the system as a whole.

Before introducing the new hierarchy of avalanches it is necessary and
worthwhile to know some feature of the new quantity, $\bar f (s)$. Firstly,
let us present some theoretical analysis. Recall the definition of $\bar f$
one can see that $\Delta \bar f (s)=\bar f (s+1)-\bar f (s)$ approaches zero
in the $L \rightarrow \infty$. An observer can not even perceived the
change of $\bar f (s)$ during the short time period since it is vanishingly small .    
However, changes at very time step are accumulated to form a relatively
distinctive change after a long time, which is perceivable for the
observers. This long time period is required to be much greater than the system size, i.e.,
$s \gg L^d$. In other words, $\bar f(s+s_0)-\bar f(s_0)$ may only be
"noticed" when $s \gg L^d$. Thus, one can not expect $\bar f(s)$ will have
great variation from the current time step to the next nearest time step,
which is vastly different from the variation of the fitness of extremal
site. The latter can vary from one value, say, $0$, to the next value, $1$
between two successive time steps. It should also be expected that there
exists an increasing tendcy of $\bar f$ versus time $s$.
This is because at each time step the least fitness is eliminated from the
system so the general fitness of the whole system will tend to increase. And
due to the slow fluctuation of the $\bar f$ the increasing behaves like a
stepwise, i.e., Devil's stepwise [2]. One then may expect to observe such
behavior, i.e., punctuated equilibrium [14], of $\bar f$ in the evolution of 
Bak-Sneppen model.

In order to show the feature of $\bar f$ versus time $s$ we performed
simulations of Bak-Sneppen model. At each time step, in addition to the
updating of the extremal sites, we also track the signals $\bar f (s)$.       
FiG. 1 shows the evolution of $\bar f (s)$ versus time during a time period
for a one-dimensional Bak-Sneppen model of size $L=200$. This plot shows
that $\bar f$ varies slightly between two successive time steps but
tends to increase in the long evolution process. Simulation of a
two-dimensional model of size $L=20$ exhibits the similar behavior of the
evolution of $\bar f (s)$. 

Before searching for the punctuated equilibrium behavior let us first
introduce another quantity, F(s), the gap of the average fitness. The
definition of F(s) is given as follows: Initial value of F(s) is equal to
$\bar f (0)$. After $s$ updates, a large $F(s) > F(0)$ opens up. The current
gap F(s) is the maximum of all $F(s^{\prime})$, for all $0 \leq s^{\prime}
< s$. FiG. 2 shows the F(s) as a step-wising increasing function of s during
the transient for a one-dimensional Bak-Sneppen model of size $L=100$.
Actually, the gap is an envelope function that tracks the increasing peaks
in $\bar f (s)$. Indeed, punctuated equilibrium behavior appears in terms of
this new quantity, $\bar f (s)$.   

By definition [14], the separate instances when the gap F(s) jumps to its
next higher value are separated by avalanches. Avalanches correspond to
plateaus in F(s) during which $\bar {f}(s) < F(s)$, which ensures the
mutation events within a single avalanche are spatially and casually
connected. A new avalanche is initiated each time the gap jumps and ends up
when the gap jumps again. As the gap increases, the probability for the
average fitness, $\bar f$, to fall below the gap increases also, and larger
and larger avalanches typically occur.

We can also obtain an exact gap equation of F(s), similar to the one found
for Bak-sneppen model in Ref. [16]. Suppose in the system the current gap 
is F(s). If F(s) is to be increased by $\Delta F$, i.e., from F(s) to
$F(s)+\Delta F$, the average number of avalanches needed is $N_{\rm
av}=\Delta F L^d /(1-F(s))$. We can guarantee $N_{\rm av} \gg 1$ by
selecting $\Delta F \gg L^{-d}$. In the large $L$ limit, $N_{\rm av}$ can be
arbitrarily large. Hence, in this limit, the average number of time steps
required to increase the gap from F(s) to $F(s)+\Delta F$ is given by the
interval $\Delta s=\langle S \rangle_{\rm F(s)} N_{\rm av}=\langle s
\rangle_{\rm F(s)} \Delta F L^d /(1-F(s))$, where $\langle S \rangle _{\rm
F(s)}$ is the average avalanche size of the plateaus in the gap function.
From the law of large numbers the fluctuation of this interval around its
average value vanishes. In the $\Delta F \rightarrow 0$ limit, $\Delta s
\rightarrow 0$. Taking the continuum limit we can obtain the differential
equation for F(s),

\begin{equation}
\frac {\rm dF(s)} {\rm ds}=\frac {1-F(s)} {L^d \langle S \rangle _{\rm
F(s)}}. 
\end{equation}

\noindent
Note this equation is exact. 

All SOC models, e.g., the BTW sandpile model [17], the earthquake models
[18], or Bak-Sneppen model [14], exhibit self-organized criticality in terms
of a power-law distribution of avalanche. It is natural to expect that we
can observe SOC in terms of the hierarchical structure of $\bar f$, which
itself manifests complexity. Using this new quantity to define the avalanche
is simply another way of observing the same phenomenon which can be observed
in other ways. As known, the emergence of complexity is independent of the
tools used to observe them provided that these tools are efficient and
strong enough. Similar to the ones used in Refs. [13,19], we present the
definition of $\bar {f}_0$-avalanche, where $\bar {f}_0$ is only a parameter 
between 0.5 and 1 to define the avalanche. Suppose at time step $s_1$, 
$\bar {f}(s_1)$ is larger than $\bar {f}_0$. If , at time step $s_1$+1, $\bar
{f}(s_1+1)$ is less than $\bar {f}_0$, this initiates a creation-annihilation
branching process. The avalanche still continues at time step $s^{\prime}$, 
if all the $\bar {f}(s)$ are less than $\bar {f}_0$ for $1 \leq s \leq 
s^{\prime}-1$. And the avalanche stops, say, at time step $s_1+S$, when $\bar {f}
(s_1+S)>\bar {f}(s_1)$. In terms of this definition, the size of the
avalanche is the number of time steps between subsequent punctuation of the
barrier $\bar {f}_0$ by the signal $\bar {f} (s)$. In the above example, the
size of the avalanche is $S$. It can be clearly seen from FiG.1 that this
definition guarantees the hierarchical structure of avalanches, larger
avalanches consists of smaller avalanches. As $\bar f_0$ is lowered, bigger
avalanches are subdivided into smaller ones. Hence, the statistics of $\bar
f_0$-avalanche will inevitably have a cutoff if $\bar f_0$ is not chosen to
be the value of $\bar f(s)$ at critical state, denoted by $\bar f_c$. We can
also define the $\bar f_c$-avalanche. Nevertheless, $\bar f_0$-avalanche in
the stationary state has the same scaling behavior as $\bar f_c$-avalanche
provided that $\bar f_0$ is close to $\bar f_c$. We measure $\bar
f_0$-avalanche distribution for one- and two-dimensional Bak-Sneppen models.
The simulation results are given in FiG. 3. The exponent $\tau$, defined by
$P(S) \sim s^{- \tau}$, is 1.80 for 1D model and 1.725 for 2D model. Another
exponent, $D$, avalanche dimension [13], defined by $n_{\rm cov} \sim
S^{D/d}$, where $n_{\rm cov}$ is the number of sites covered by an
avalanche, and $d$ is the space dimension, is measured. We find $D$=2.45
for 1D model and 1.55 for 2D model.          
  
Up to now, a question is still unsolved. It is about the critical value of
$\bar f$, $\bar f_c$. This may be a hard bone if the system size is finite,
but when the consider the L limit, everything will be smooth and can be
easily accomplished. Recall the evolution of Bak-Sneppen model, or the
detailed research of this model [13], the densities of sites with random
numbers is uniform above G [13] and vanishes below G in the $L \rightarrow
\infty$, where G is the gap of extremal site and detailed information of
it can be found, for instance, in Ref. [13]. Hence, one can obtain, 

\begin{equation}
\lim_{L \rightarrow \infty}\bar f (s)=
\lim_{L \rightarrow \infty}\frac {1+G(s)} {2} .
\end{equation}

\noindent
Interestingly,
inserting Eq. (3) into the gap equation of G found in Ref. [19], one can
immediately obtain Eq. (2). Please note that Eq. (2) is also valid for
finite-size systems. From Eq. (3) one can immediately obtain,
\begin{equation}
\lim_{L \rightarrow \infty} \bar {f}_c=\lim_{L \rightarrow \infty}\frac {1+f_c} {2} .
\end{equation}
\noindent
Hence, $\bar {f}_c$ can be easily determined from Eq. (4). Using the results
of $f_c$ provided by Refs. [13,20], one can obtain $\bar {f}_c$, 0.83351 for
1D model and 0.66443 for 2D model. However, Eqs. (3) and (4) are not valid
for a finite-size system, since one can not ensure the distribution of
random numbers during a
finite-size system is really uniform. Due to the fluctuation of $\bar {f}(s)$ 
it is extremely difficult for one to determine exactly the critical value of $\bar {f}$ 
for a finite-size system. One may estimate $\bar {f}_c$ for a finite-size
system using the simulation. We find that this value weakly depends on the system
size. When the system size is very large $\bar f_c$ approaches the
corresponding value for infinite systems. Actually, the value of $\bar f_c$
itself is not so important. FiG. 4 shows the fluctuation of $\bar f$ for a
one-dimensional model of size $L=200$ near its critical state. We note, in
this figure, $\bar f$ fluctuates slightly around some average  value and does not tend
to increase any more for a long time. We may say that the system approaches
its stationary state. In this sense, we suggest that $\bar f$ may be a good
quantity in determining the emergence of criticality. That is, the great
fluctuation of $f_{\rm min}$ will not affect us to determine when we approach
the critical state. We need only to know the feature of $\bar f$. This is more
reasonable and easily accepted since $\bar f$ is a global quantity and
condenses information of the system and its components.   
    
Why we call the $\bar f_0$-avalanche a new hierarchy of avalanches? Firstly,
this kind of avalanche is defined on the global level, in terms of the new
global quantity, $\bar f$. The background of this definition is different
from any one used before. This new kind of avalanche reflects the fractal
geometry in terms of the global feature. Secondly, one can notice that the
exponents $\tau$ obtained in our simulation are different from the ones found in Ref. [13].
From this point of view, one can judge that this kind of avalanche is
totally different from any one observed before. Hence, its a new kind of
avalanche. 

Self-organized criticality is suggested by Bak et al to be the "fingerprint"
of a large variety of complex system (they call system with variability as
complex) and is represented by a scale-free line on a log-log plot. In order
to know the criticality of a system one needs to know when the system
reaches the stable stationary state where the phase transition occurs. It is
extremely difficult and almost impossible for one to know when a system in nature
approaches, not even reaches , its critical state. One can just study the
ubiquitous fractal geometrical structure carved by avalanches through
thousands of millions of years. However, in laboratory experiments and
computer simulations, one needs a criterion to judge when stationary state
approaches, even, reaches, since statistics of avalanches may only be done
under critical state of the system. Given Bak-Sneppen model, when the
extremal signal, $f_{\rm min}$, approaches to the self-organized threshold,
$f_c$, the ecosystem reaches its stationary state. However, $f_{\rm min}$
itself fluctuates greatly time to time, which brings great difficulty in
determining the appearance of criticality. Thus, we provide a new quantity,
$\bar f$, for a candidate in judging the emergence of criticality. As shown,
$\bar f$ is relatively stable in a short time period. Hence, when $\bar f$
does not tend to increase any more, we may say that the system approaches
its stationary state. And, we can observe criticality in a rather long time
period. Surely, the emergence of criticality is rather complex, other
physical mechanism is needed, this is what we will consider in the future
work.

In conclusion, a new hierarchy of avalanches is observed in Bak-Sneppen
model. A new quantity, $\bar f$, is presented and is suggested by us to be a
possible candidate in determining the emergence of criticality. An exact gap
equation and simulation results are also given.

This work was supported by NSFC in China. We thank Prof. T. Meng for
correspondence and helpful discussions.

\vskip 0.5cm
\begin{center}
\bf {Figure Captions}
\end{center}

\vskip 0.2cm
FiG. 1: The variation of $\bar f$ versus time during a time period for a (a)
        one-dimensional Bak-Sneppen model of size $L=200$ and 
	(b) two-dimensional Bak-Sneppen model of size $L=20$. The plots 
	show the hierarchical structure of $\bar f$. 

FiG. 2: Punctuated equilibrium of $\bar f$ for a (a) one-dimensional
	Bak-Sneppen model of size $L=200$ and (b) two-dimensional 
	Bak-Sneppen model of size $L=20$. We track the increasing signal of 
	$\bar f_s$, i.e, F(s). 

FiG. 3: Distribution of $\bar f_0$-avalanche for a (a) one-dimensional
	Bak-Sneppen model of size $L=200$ and (b) two-dimensional
	Bak-Sneppen model of size $L=20$. $\bar f_0$ for (a) is chosen 
	to be 0.821, and for (b), 0.648. The slopes are -1.800 and
	 -1.725 for the two plots respectively.

FiG. 4: The fluctuation of $\bar f$ around the critical state of a
	one-dimensional Bak-Sneppen model of size $L=200$.
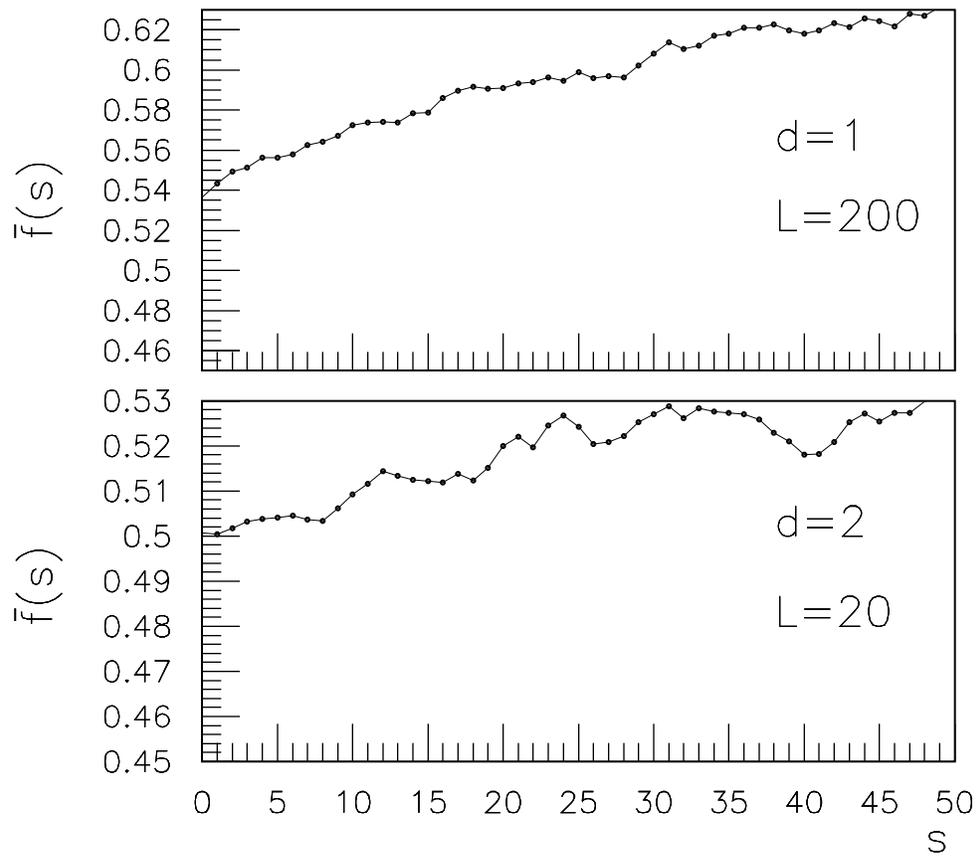
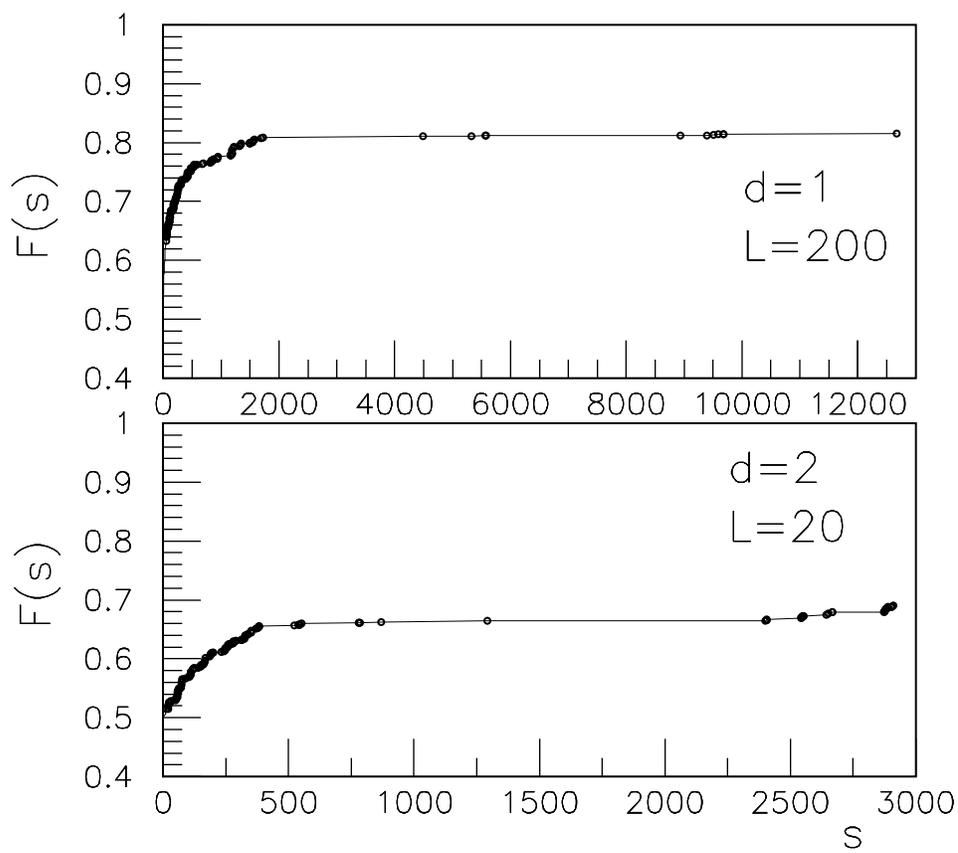
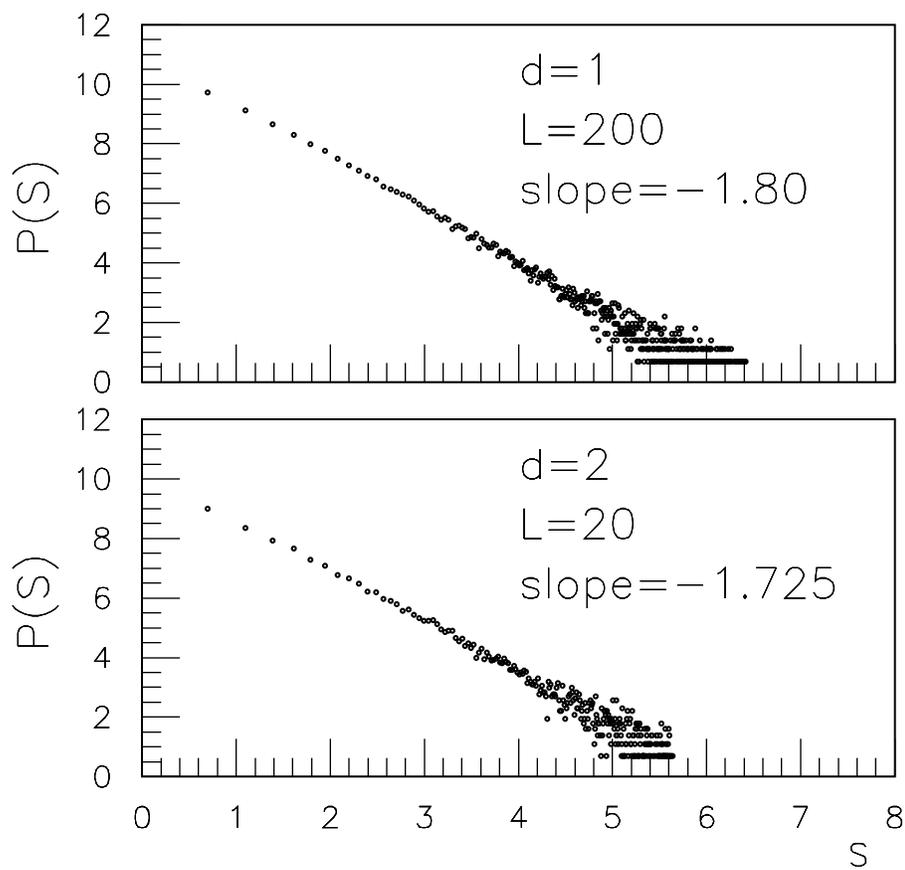
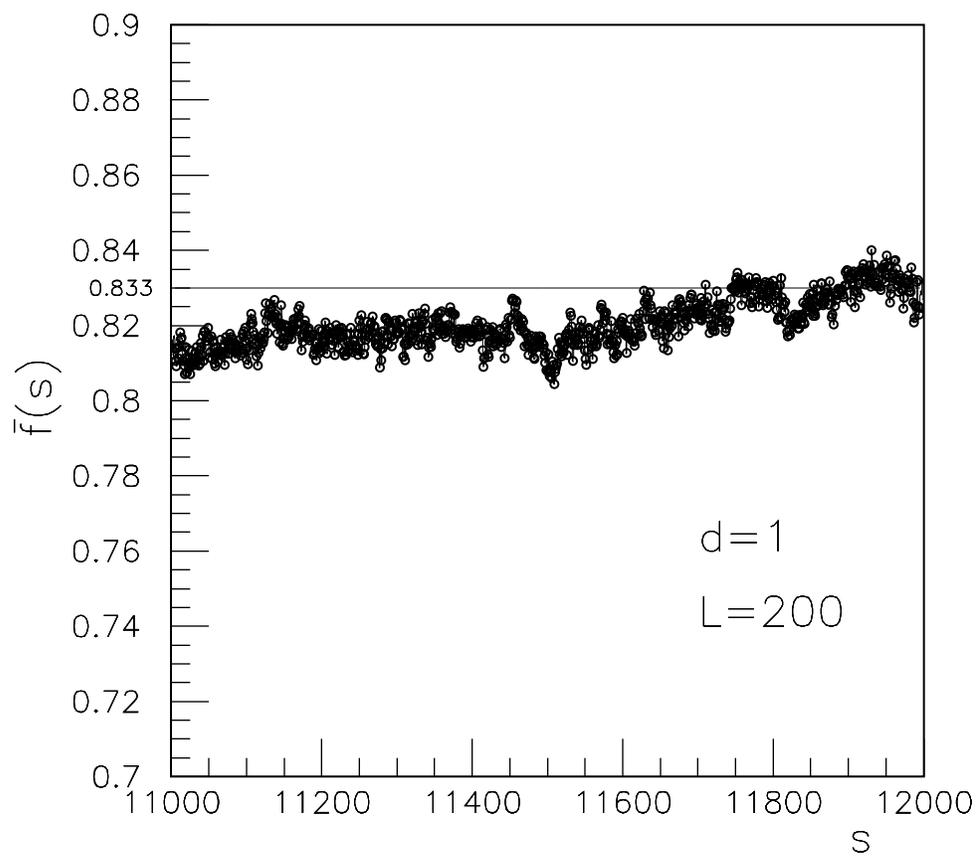
\end{document}